\documentclass[aps,prl,twocolumn,groupedaddress,showpacs]{revtex4}

\usepackage{CJK}

\usepackage{graphicx}

\usepackage{dcolumn}

\usepackage{bm}
\usepackage[normalem]{ulem}
\usepackage{color}

\usepackage{epstopdf}

\newcommand{\xto}{x\rightarrow 0}

\newcommand{\bq}{\mathbf{q}}
\newcommand{\bv}{\mathbf{v}}

\newcommand{\br}{\mathbf{r}}

\newcommand{\bM}{\mathbf{M}}
\newcommand{\bff}{\mathbf{f}}
\newcommand{\bu}{\mathbf{u}}

\newcommand{\sep}{ \ \ \ , \ \ \ }

\newcommand{\beq}{\begin{equation}}
\newcommand{\eeq}{\end{equation}}
\newcommand{\beqn}{\begin{eqnarray}}
\newcommand{\eeqn}{\end{eqnarray}}
\newcommand{\pp}{\partial}
\newcommand{\dd}{{\rm d}}
\newcommand{\ee}{{\rm e}}
\newcommand{\eq}{Eq.\ }

\newcommand{\cO}{{\cal O}}

\newcommand{\la}{\langle}

\newcommand{\ra}{\rangle}

\begin{document}

\begin{CJK*}{GBK}{}

\title{Surprising mappings of 2D polar active fluids to 2D soap and 1D sandblasting
}

\author{Leiming Chen}
\address{College of Science, China University of Mining and Technology, Xuzhou Jiangsu, 221116, P. R. China}
\author{Chiu Fan Lee}
\address{Department of Bioengineering, Imperial College London, South Kensington Campus, London SW7 2AZ, U.K.}
\author{John Toner}
\address{Department of Physics and Institute of Theoretical
Science, University of Oregon, Eugene, OR $97403$}
\address{Max Planck Institute for the Physics of Complex Systems, N\"othnitzer Str. 38, 01187 Dresden, Germany}

\begin{abstract}
Active fluids and growing interfaces are two well-studied but very different non-equilibrium systems. 
Each exhibits  non-equilibrium behavior quite different from that of their equilibrium counterparts.
Here we demonstrate a surprising connection between these two:
the ordered phase of incompressible polar active fluids 
in two spatial dimensions without
momentum conservation,
and growing one-dimensional interfaces (that is,
the 1+1-dimensional  Kardar-Parisi-Zhang equation), in fact
belong to the same universality class. This
universality class also includes two equilibrium systems : two-dimensional smectic liquid crystals, and a
peculiar kind of constrained two-dimensional ferromagnet.
We use these connections to show that two-dimensional
incompressible flocks are robust against
fluctuations, and exhibit universal long-ranged,
anisotropic spatio-temporal correlations of
those fluctuations.
We also thereby determine the exact values of the
anisotropy exponent $\zeta$ and the roughness exponents
$\chi_{_{x,y}}$ that characterize these correlations.

\end{abstract}
\maketitle
\end{CJK*}

\noindent{\bf Introduction} .
\newline
Non-equilibrium  systems can behave
radically differently from their equilibrium counterparts.  Two of the most striking
examples of such exotic non-equilibrium behavior are moving interfaces
(e.g., the surface of a growing crystal) \cite{surface}, and ``flocks"
(i.e., coherently moving states of polar active fluids) \cite{boids1, boids2, boids3, boids4,Vicsek1, Vicsek2}.
The former is described by the Kardar-Parisi-Zhang (KPZ) equation \cite{KPZ}, which is also a model for erosion (i.e., sandblasting). This equation predicts
that a two-dimensional (2D) moving interface (i.e., the surface of a three-dimensional crystal) is far rougher than the surface of a crystal in equilibrium.
In contrast, hydrodynamic theories of polar active fluids \cite{TT1, TT2, TT3, TT4,TT5} predict that a large
collection of ``active" (i.e., non-equilibrium) moving particles  (which could be
anything from motile organisms to molecular motor propelled biological
macromolecules \cite{boids1, boids2, boids3, boids4, Vicsek1, Vicsek2, TT1, TT2,TT3,TT4, TT5, dictyo1, dictyo2, rappel1, microtub})
can develop long-ranged orientational order in 
2D, while their equilibrium counterparts
(e.g., ferromagnets),
by the Mermin-Wagner \cite{MW1,MW2} theorem, cannot.
At the same time,  many non-equilibrium systems can also be mapped onto 
equilibrium systems \cite{wioland}; an example of this that proves very relevant is the connection between the 1+1-dimension KPZ model and the defect-free 2D smectic (i.e., soap) model \cite{Golubo1, Golubo2}. Here, we add a living system to this list by showing that  generic incompressible active polar fluids, e.g., an incompressible bird flock,  all belong to the same universality class.

Since many  fluids  flow   much slower than the speed of sound, a great deal of the work done over the past two centuries on equilibrium
fluids has focused on incompressible fluids \cite{Landau,FNS}.  In this paper, we
consider 2D active incompressible fluids; more specifically,
we consider them in rotation invariant, but non-Galilean-invariant situations in
which momentum is not conserved (e.g., active fluids moving over an isotropic
frictional substrate such as
cells crawling on a substrate).
Such an active system contains rich physics: it has recently been shown that their
static-moving transition belongs to a new universality class \cite{chen_njp15}.
Here, we focus on the long-range properties of the system in the moving phase.

We note that the incompressibility condition
is not merely a theoretical contrivance; not only can it be readily
simulated \cite{wensink_pnas12, ramaswamy_jcomputphys15}
but it can arise in a variety of real experimental situations,  including 
 systems with long-ranged repulsive interactions \cite{turner}, and dense systems of active particles with strong repulsive short-ranged interactions, such as bacteria  \cite{wensink_pnas12}. 
In addition, incompressibility plays an important role in the  motile colloidal systems in fluid-filled microfluidic channels recently studied \cite{bartolo},  although these systems differ in detail from those we study
 here in being two component (background fluid plus colloids).

\begin{figure}
\begin{center}
\includegraphics[scale=.38]{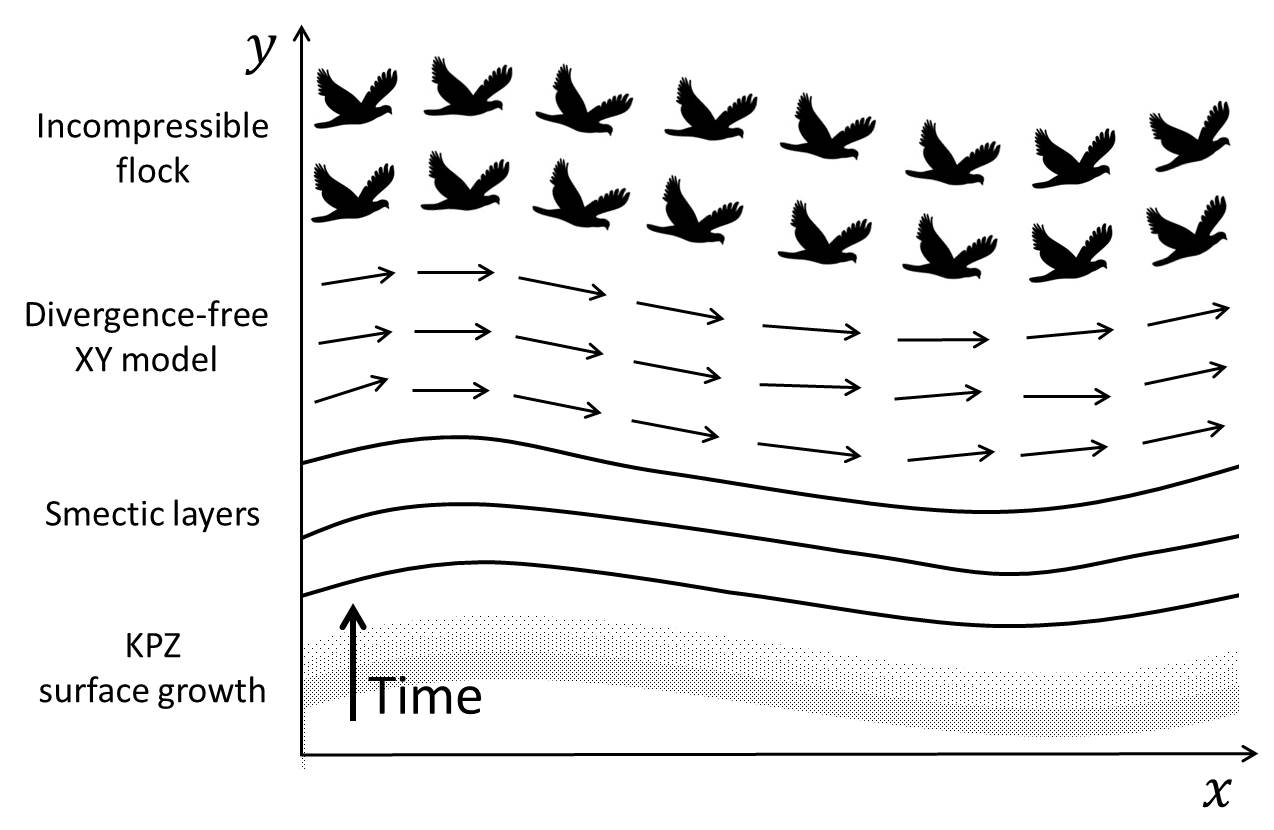}
\end{center}
\caption{ $|$ {\bf  Visual representation of the mappings.}
The flow
lines of  the ordered phase of a 2D incompressible polar active fluid, the magnetization lines of the ordered phase of divergence-free 2D $XY$
magnets, dislocation-free 2D smectic layers, and the surfaces of a growing one-dimensional  crystal  (which can be obtained by taking equal-time-interval snap shots), undulate in
exactly the same way over space; their fluctuations share exactly the same asymptotic scaling
behavior at large length scales. 
Note that the vertical axis is time for KPZ surface growth and  the $y$ Cartesian coordinate  for the other three systems.
}
\label{fig1}
\end{figure}

In this paper, we formulate a hydrodynamic (i.e., long-wavelength and long-time) theory of the ordered, 
moving phase of a 2D incompressible polar active fluid.
We find that the equal-time velocity correlation functions of the type of incompressible polar active fluids 
we study here can be mapped exactly onto those of two equilibrium problems: 
a divergence-free 2D $XY$ model  (a peculiar type of ferromagnet
 different from  ordinary 
ferromagnets, which are divergenceful) 
and a dislocation-free 2D
smectic A liquid crystal \cite{deGennes, first, Golubo1,Golubo2, Kashuba}, as well as onto  the time dependent correlation functions of the non-equilibrium 
1+1-dimensional KPZ equation \cite{KPZ}.  The mapping of the 2D smectic onto the $1+1$-dimensional KPZ equation was discovered by Golubovic and Wang \cite{Golubo1, Golubo2}; the other two mappings are new (although 2D ferromagnets with 2D dipolar interactions,  which are  similar but not identical systems, have also been mapped onto 2D smectics \cite{Kashuba}).
This series of mappings is illustrated 
in Fig. \ref{fig1}. 

Our results imply in particular  
that incompressible polar active fluids can develop long-ranged orientational order (by developing a non-zero 
mean velocity $\la{\bf v}\ra$) in two dimensions,
 just as found previously for compressible polar active fluids, but in complete contrast to  
their equilibrium counterparts (i.e., ordinary 
 divergenceful ferromagnets) with underlying rotation invariance, which  cannot so order.  However, the scaling behavior of the 
 velocity correlation functions is very different from those for compressible 
 polar active fluids studied in Ref. \cite{TT3, TT4}.
Specifically, we find 
that the 
equal-time 
velocity
correlation  function in the ordered phase 
has the following limiting behaviours:
\begin{eqnarray}
&& 
\left<|\bv(\br,t)-\bv(\br',t)|^2\right> \nonumber\\
&=&\left\{
\begin{array}{ll}
C_0-AY^{-2/3} \ , & \kappa\ll 1
\\
C_0-{9\over 2}c^2{A\over X} e^{-\Phi(\kappa)}\left[1+{4\over 9}\left({x-x'\over y-y'}\right)^2\right] \ , & \kappa\gg 1
\end{array}~~~
\right.\label{LC_scaling}
\end{eqnarray}
\noindent where $X\equiv |x-x'|/\xi_x$ and $Y\equiv|y-y'|/\xi_y$ are rescaled lengths in the  $x$ and $y$ 
directions, and we define the scaling 
ratio $\kappa\equiv{X\over Y^{2/3}}$. Here  the function $\Phi(\kappa\gg 1)\approx c\kappa^3$ and the constant $c\approx 0.295$ are both universal (i.e., system-independent), while
   $C_0$ and $A$ are non-universal (i.e., system-dependent), positive, finite constants, and $\xi_{x,y}$ are non-universal  lengths.  
%
Note that  the fact that $\left<|\bv(\br,t)-\bv(\br',t)|^2\right>$ goes to a finite value in the  large separation  limit $|\br-\br'|\to\infty$ implies
long-ranged  orientational order.
\newline

\noindent{\bf Results}
\newline
{\bf Model.}
We start with the hydrodynamic model for compressible 
polar active fluids  without momentum conservation \cite{TT1, TT3, TT4}:
\begin{eqnarray}
\partial_t\rho +\nabla\cdot(\bv\rho)=0
\label{conservation}
\end{eqnarray}

\begin{widetext}\begin{eqnarray}\partial_{t}\bv+\lambda_1(\bv\cdot\nabla)\bv+\lambda_2(\nabla\cdot\bv)\bv+\lambda_3 \nabla(|\bv|^2)=U\bv -\nabla P -\bv\left( \bv \cdot \nabla P_2 \right) +\mu_{{\rm B}} \nabla(\nabla\cdot \bv)+ \mu_{\rm T}\nabla^{2}\bv +\mu_{2}(\bv\cdot\nabla)^{2}\bv+\mathbf{f}\nonumber \\\label{EOM}\end{eqnarray}\end{widetext}
where $\bv(\br,t)$, and $\rho(\br,t)$ are respectively the coarse grained continuous velocity and density fields.
All of the parameters $\lambda_i (i = 1 \to 3)$,
$U$, the ``damping coefficients" $\mu_{\rm B,T,2}$, the  ``isotropic pressure'' $P(\rho,
v)$ and the  ``anisotropic Pressure'' $P_2 (\rho, v)$
are, in general, functions of the density $\rho$ and the
magnitude $v\equiv|\bv|$ of the local velocity.
Note that we omit higher order damping terms because, as our analysis 
will show later, they are irrelevant. In 
addition,  because we focus here on the ordered phase, $\mu_{{\rm T},{\rm B},2}$ is 
taken to be positive, as required for the stability of the ordered phase.

The $U$ term makes the local
$\bv$ have a nonzero magnitude $v_0$
in the ordered phase, by the simple expedient of having $U>0$ for $v<v_0$,
$U=0$ for $v=v_0$, and $U<0$ for $v>v_0$. 
The $\mathbf{f}$ term is a random
driving force. It is  assumed to be Gaussian with
white noise correlations:
\begin{eqnarray}
\label{eq:noise}
   \la f_{i}(\br,t)f_{j}(\br',t')\ra=2D
\delta_{ij}\delta^{d}(\br-\br')\delta(t-t')
\label{white noise}
\end{eqnarray}
where the ``noise strength" $D$ is a constant parameter of the system, and $i,j$ denote
Cartesian components. Note that 
in contrast to thermal fluids (e.g., Model A in \cite{FNS}), we are concerned with active systems that are not momentum conserving. As a result, the leading contribution to the 
noise correlations is of the form depicted in (\ref{eq:noise}).

We now take the incompressible limit by taking  the isotropic pressure $P$ only
to be extremely sensitive to departures from the mean density $\rho_0$. One could alternatively consider making $U(\rho, v)$ and $P_2(\rho, v)$
extremely sensitive to changes in $\rho$ as well. This would be appropriate for an
active fluid near its ``active jamming" \cite{Marchetti} transition, since in that regime
a small change in the local density can change the speed from a non-zero value
for $\rho<\rho_{\rm jam}$ to zero for $\rho>\rho_{\rm jam}$.
We will discuss this case in a future publication.

Focusing here on the case in which only the isotropic pressure $P$ becomes
extremely sensitive to changes in the density, we see that, in this limit, in which the
isotropic pressure 
suppresses density fluctuations extremely effectively, changes in the density are 
too small to affect  $U(\rho, v)$, $\lambda_{1,2,3}(\rho, v)$, $\mu_{{\rm B},{\rm T},2}(\rho, v)$,
and $P_2(\rho, v)$.  As a result, in the incompressible limit taken this way,
all of them effectively become functions only of the
speed $v$; their $\rho$-dependence drops out since $\rho$ is
essentially constant.

Another consequence of the suppression of density fluctuations by the isotropic pressure $P$ is that the continuity equation (\ref{conservation}) reduces 
to the familiar condition 
for incompressible flow, 
\begin{eqnarray}
\nabla\cdot\bv=0 \,,
\label{inc}
\end{eqnarray}
which can, as in simple fluid mechanics, be used to determine the isotropic pressure $P$.

All of the above discussion taken together leads to the following equation of motion in tensor notation 
for an incompressible polar active fluid, ignoring irrelevant terms:
\begin{widetext}
\beqn
\pp_t v_m = -\pp_m P +U(v) v_m - \lambda_1(v)v_n(\pp_n v_m)
- \lambda_4(v)v_mv_nv_{\ell}(\pp_n v_{\ell})
 +\mu_{\rm T}(v) \pp_n \pp_n v_m
+  \mu_2 (v)v_\ell v_n\pp_\ell \pp_n v_m
+ f_m
\ ,\label{EOM1}
\eeqn
\end{widetext}
\noindent
where $\lambda_4(v)\equiv  {1\over v}{dP_2(v)\over dv}$, and the $\lambda_2$ and
$\mu_B$ terms vanish due to the incompressibility constraint  $\nabla\cdot\bv=0$ on $\bv$. 
In writing (\ref{EOM1}), we absorb a term $W(v)$ into the pressure $P$, where $W(v)$ is derived from $\lambda_3(v)$ by solving $ {1\over 2v}{dW\over dv}=\lambda_3(v)$.

We now analyze the implications of equation (\ref{EOM1}) for the ordered state.
\newline

\noindent{\bf Linear theory.}
In the ordered phase, the system spontaneously breaks rotational symmetry by moving
on average along some spontaneously chosen direction which we call $\hat{x}$;
we call the direction orthogonal to this $\hat{y}$.
In the absence of fluctuations (i.e., if we set the noise $\bff$ in (\ref{EOM1}) to zero),
the velocity will be the same everywhere in
space and time, and have magnitude
$v_0$, which we remind
the reader is defined by
$U(v_0)=0$.
We treat
 fluctuations by expanding
$\bv$ around $v_0\hat{x}$, defining  $\bu(\br, t)$ as the small fluctuation in the velocity field about this mean:
\beqn
\bv =(v_0+u_x(\br, t))\hat{x} + u_y(\br, t) \hat{y}
\ .\label{Expension1}
\eeqn

Plugging Eq. (\ref{Expension1}) into Eq. (\ref{EOM1}) and
expanding to linear order in
$\bu$, 
leads to a linear stochastic partial differential equation with constant coefficients. Like all such equations, this can be solved simply by spatio-temporally Fourier transforming, and solving the resultant linear algebraic equations for the Fourier transformed field $\bu(\bq,\omega)$ in terms of the Fourier transformed noise $\bff(\bq,\omega)$. We can thereby relate the two point correlation function $\la |u_y(\bq, \omega)|^2\ra$ to the known correlations (\ref{white noise}) of the
random force $\bff$. Integrating the result over all frequencies $\omega$,  and dividing by $2\pi$, gives the equal time, spatially Fourier transformed velocity autocorrelation $ \langle  |u_y(\bq, t)|^2\rangle$. Details of this straightforward calculation are given in ``Methods"; the result is
\begin{eqnarray}
 \langle  |u_y(\bq, t)|^2\rangle={Dq_x^2\over 2\alpha q_y^2+\Gamma (\bq)q^2}\approx{Dq_x^2\over 2\alpha q_y^2+\mu q_x^4}
 \,,
 \label{Uu3}
\end{eqnarray}
where $\Gamma(\bq)\equiv\mu q_x^2+\mu_{\rm T}^0q_y^2$
with $\mu\equiv \mu_{\rm T}^0+\mu_2^0v_0^2$, where $\mu_{{\rm T},2}^0$ are $\mu_{\rm T,2}(v)$ evaluated at $v=v_0$, and
the second, approximate equality applies for all $\bq\rightarrow \bf{0}$. This can be seen by noting that, for $q_y\gg q_x^2$ and $\bq\rightarrow \bf{0}$, $q_y^2\gg \Gamma(\bq)q^2$, while for $q_y\lesssim q_x^2$ and $\bq\rightarrow \bf{0}$,  $\Gamma(\bq)q^2\approx \mu q_x^4$. Hence, in both cases, (which together cover all possible ranges of $\bq$ for  $\bq\rightarrow \bf{0}$), the approximation $2\alpha q_y^2+\Gamma (\bq)q^2\approx  2\alpha q_y^2+\mu q_x^4$ is valid.

We can now obtain the real space transverse fluctuations 
\beqn
\la u_y^2(\br, t) \ra=\int_{{q_x}\gtrsim{1\over L}}{d^2q\over (2\pi)^2} \langle|u_y(\bq, t)|^2\rangle \,,
\label{realspacefluc}
\eeqn
where $L$ is the lateral extent of the system in the $x$-direction (its extent in the $y$-direction is taken for the purposes of this argument to be infinite).  Note that the longitudinal fluctuations $\la u_x^2(\br, t)\ra$ are negligibale compared to $\la u_y^2(\br, t)\ra$.  Using (\ref{Uu3}), the integral in (\ref{realspacefluc}) is readily seen to converge in the infra-red, and, hence, as system size $L\rightarrow\infty$. Since the integral is finite, and proportional to the noise strength $D$, it is clear that, for sufficiently small $D$, the transverse fluctuations  $\la u_y^2(\br, t)\ra$ can be made small enough that long-ranged orientational order
-  i.e., a non-zero $\langle\bv(\br,t)\rangle$ - is preserved in the presence of fluctuations; therefore, the ordered state is stable against fluctuations for sufficiently small noise strength $D$.

We show in the next section that this conclusion remains valid when nonlinear effects are taken into
account (even though those nonlinearities change the scaling laws from those predicted
by the linear theory).
\newline

\noindent{\bf Nonlinear Theory.}
We begin by expanding the full equation of motion (\ref{EOM1}) to higher order in 
$\bu$.
This gives
\begin{widetext}
\begin{eqnarray}
\pp_t u_m &=& -\pp_m P -2\alpha u_x\delta_{mx} - \lambda^0_1v_0\pp_x u_m
 +\mu^0_{\rm T} \nabla^2 u_m
+  \mu_2^0v_0^2 \pp_x^2u_m+f_m\nonumber
\\
&&-{\alpha\over v_0}\left({u_y^3\over v_0}\delta_{my}+2u_xu_y\delta_{my}+u_y^2\delta_{mx}\right)
-\lambda^0_1u_y\pp_yu_{y}\delta_{my}
\ ,\label{EOMNL1}
\end{eqnarray}
\end{widetext}
where the superscript ``0'' means that the $v$-dependent coefficients are
evaluated at $v=v_0$, and we define the ``longitudinal mass"
$\alpha\equiv -{v_0\over 2}\left({dU(v)\over dv}\right)_{v=v_0}$.

The first line of equation (\ref{EOMNL1}) contains the linear terms, including the noise ${\bf f}$; the first three terms on the second line 
are the relevant non-linearities, while the fourth term proves to be irrelevant, as we'll soon show.

In writing (\ref{EOMNL1}), we have neglected ``obviously irrelevant" terms, by which we mean terms that differ from those explicitly displayed in (\ref{EOMNL1}) by having more powers of the small fluctuations $\bu$, or more spatial derivatives of a given type. For more discussion of these ``obviously irrelevant"  terms, see  ``Methods".  Note that only one of the non-linearities associated with the $\lambda_{1,2,3}$ terms, namely, $\lambda^0_1u_y\pp_yu_{y}$ actually remains at this point.

To proceed further, we must power count more carefully.

We only need to calculate one of the two fields $u_{x,y}$, since they are related by the incompressibility condition $\nabla\cdot\bv=0$. We 
choose to solve for $u_y$; its Fourier transformed equation of motion can be obtained by Fourier transforming (\ref{EOMNL1}) and acting on both sides  of the resultant equation with the transverse projection operator $P_{lm}(\bq)=\delta_{lm}-q_lq_m/q^2$ which projects orthogonal to the spatial wavevector $\bq$. This eliminates the pressure term. Taking the $l=y$
component of the resulting equation gives: 

\begin{widetext}
\begin{eqnarray}
\partial_t u_y(\bq,t)&=& -iv_1q_xu_y(\bq,t)-\Gamma(\bq)u_y(\bq,t)+P_{yx}(\bq)\mathcal{F}_{\bq}\left[-2\alpha
                        \left(u_x(\br,t)+{u_y^2(\br,t)\over 2v_0}\right)
                        \right]\nonumber\\
&&+P_{yy}(\bq)\mathcal{F}_{\bq}\left[-{\alpha\over v_0}\left({u_y^3\over v_0}+2u_xu_y\right)-\lambda_1^0u_y\pp_yu_y\right]+P_{ym}(\bq)f_m(\bq,t)\,,
\label{FullEOM1}
\end{eqnarray}
\end{widetext}
where $\mathcal{F}_{\bq}$ represents the 
Fourier component at wavevector $\bq$, i.e.,   $\mathcal{F}_{\bq}[g(\br)] \equiv \int d^2 r\, g(\br) e^{-i \bq \cdot \br}$; the ``bare" value of the speed $v_1$, before rescaling and renormalization, is $v_1=\lambda^0_1v_0$,  and $\Gamma(\bq)$ is given after equation (\ref{Uu3}).

We now rescale co-ordinates
($x,y$), time $t$, and the components of the real space velocity field $u_{x,y}(\br,t)$ according to 
\beqn
&&x \mapsto  e^{\ell} x \,,~
y \mapsto  e^{\zeta\ell} y \,,~
t \mapsto  e^{z\ell} t
\label{2dResclx}\\
&&u_y(\br,t)\mapsto  e^{\chi_y\ell} u_y(\br,t) \,,\\
&&u_x(\br,t)\mapsto  e^{\chi_x\ell} u_x(\br,t) =e^{\left(\chi_y+1-\zeta\right)\ell} u_x(\br,t) \,,
\label{2dResclux}
\eeqn
where the scalings of $u_x(\br,t)$ and $u_y(\br,t)$ are related by the
incompressibility condition. Note that our convention for the anisotropy exponent here is exactly the opposite of that used in references \cite{TT1, TT2, TT3,TT4,TT5}; that is, we define $\zeta$ by $q_y\sim q_x^\zeta$ being the dominant regime of wavevector, while \cite{TT1, TT2, TT3,TT4,TT5} defines this regime as $q_x\sim q_y^\zeta$.

Upon this rescaling,  the form of \eq(\ref{FullEOM1}) remains
unchanged, but the various coefficients become dependent on the rescaling parameter $\ell$.

Details of this simple power counting (including the slightly subtle question of how to rescale the projection operators) are given in ``Methods". The results for the three parameters (damping coefficient $\mu$, ``longitudinal mass" $\alpha$, and noise strength $D$) that control the size of the fluctuations in the linear theory are: $\mu\mapsto  e^{\left(z-2\right)\ell} \mu$, $\alpha \mapsto  e^{\left(z-2\zeta+2\right)\ell} \alpha$, and $D \mapsto  e^{\left(z-2\chi_y-\zeta-1\right)\ell} D$.

We now use the standard renormalization group logic to assess the importance of the non-linear terms in (\ref{FullEOM1}). This logic is to choose  the rescaling exponents $z$, $\zeta$, and $\chi_y$ so as to keep the size of the fluctuations in the field $\bu$ fixed upon rescaling. This is clearly accomplished by keeping  $\alpha$,  $\mu$, and $D$ fixed. From the rescalings just found, this leads to three simple linear
equations in the three unknown exponents $z$, $\zeta$, and $\chi_y$; solving these, we find the values of these exponents in the linearized theory:
$\zeta_{_{\rm lin}}=z_{_{\rm lin}}=2, \chi_{_{y\rm lin}}=-1$.
With these exponents in hand, we can now assess the importance of the non-linear terms in (\ref{FullEOM1}) at long length scales, simply by looking at how their coefficients rescale. (We don't have to worry about the size of the actual non-linear terms themselves changing upon rescaling, because we have chosen the rescalings 
to keep them constant 
in the linear theory.) We find that all of the non-linearities whose coefficients are proportional to $\alpha$ are ``relevant" (i.e., grow upon rescaling), while those associated with the last remaining non-linearity, $\lambda^0_1$, associated with the $\lambda$ terms get smaller upon rescaling: $\lambda^0_1 \mapsto  e^{-{\ell\over 2}} \lambda^0_1$. Hence, this term will not affect the long-distance behavior, and can be dropped from the problem. This is very different from the compressible problem, in which the $\alpha$ non-linearities are unimportant, while the $\lambda$ ones dominate; the reasons for this difference are discussed in ``Methods".

Dropping the $\lambda_1^0$ term in (\ref{EOMNL1}), and making a Galilean transformation to a ``pseudo-co-moving" co-ordinate system moving in the direction $\hat{x}$ of mean flock motion at speed $v_1 \equiv \lambda^0_1v_0$ 
to eliminate the ``convective term" $v_1\pp_xu_m$
from the right hand side of (\ref{EOMNL1}), 
leaves us with our final simplified form for the equation of motion:
\begin{eqnarray}
\pp_t u_m =&& -\pp_m P -2\alpha \left(u_x+{u_y^2\over 2v_0}\right)\delta_{xm} \nonumber\\
&&-{2\alpha\over v_0} \left(u_x+{u_y^2\over 2v_0}\right)u_y\delta_{ym} \nonumber\\
&&+  \mu \pp_x^2u_m+\mu_{\rm T}^0\pp_y^2u_m+f_m
\ .\label{EOMNL3}
\end{eqnarray}

We now  show 
that  Eq.~(\ref{EOMNL3})  also describes an equilibrium system: the ordered phase of the 
2D $XY$ model subject to the  divergence-free constraint
$\mathbf{\nabla}\cdot\bM=0$, where $\bM$ is
the magnetization. This connection enables us to use purely equilibrium statistical mechanics (in particular, the Boltzmann distribution) to determine the equal-time correlations of 2D incompressible polar active fluids.
\newline

\noindent {\bf Divergence-free 2D $XY$ model.} 
The 2D $XY$ model describes a
2D ferromagnet whose magnetization field $\mathbf M(\br)$ and position $\br$ both have 
two components.
The
Hamiltonian for this model can be written, ignoring irrelevant terms,  as \cite{Lubensky}
\beq
H_{\rm _{XY}}=\int \dd^2r \left[V(|\mathbf M|)+\frac{1}{2} \mu|\vec{\nabla} \mathbf M|^2\right]\, ,\label{Hxy1}
\eeq
where $\mu$ is the ``spin wave stiffness".
In the ordered phase, the ``potential" $V(|\mathbf M|)$ has a circle of global minima at a non-zero value of $|\mathbf M|$, which we will take to be $v_0$.

Expanding in small fluctuations about this minimum by writing $\mathbf M=\left(v_0+u_x\right)\hat{x}+u_y\hat{y}$,
we obtain, keeping only ``relevant" terms,
\beq
H_{\rm _{XY}}=\frac{1}{2} \int \dd^2r \left[2\alpha\left(u_x+{u_y^2\over 2v_0}\right)^2+\mu|\vec{\nabla} \bu|^2\right]\, ,\label{Hxy2}
\eeq
where we define the ``longitudinal mass''
$2\alpha\equiv \left.\left(\partial^2 V\over\partial |\bM|^2\right)\right|_{|\bM|=v_0}$.

We now add to this model
the divergence-free constraint $\mathbf {\nabla}\cdot \mathbf M=0$, which obviously implies
$\mathbf {\nabla}\cdot \mathbf u=0$. To
enforce this constraint, we introduce to the Hamiltonian a Lagrange multiplier $P(\br)$:
\beq
H'=H_{\rm _{XY}}- \int \dd^2r \ P(\br)(\mathbf {\nabla}\cdot \mathbf u)\, .
\label{Hconst}
\eeq

The simplest dynamical model that relaxes back to the equilibrium Boltzmann distribution $e^{-\beta H'(\bu)}$ for the Hamiltonian $H'$ is \cite{Ma, Lubensky} the ``time-dependent-Ginsburg-Landau" (TDGL) model
$ \pp_t u_l=-{\delta H'/ \delta u_l}+f_l$, where  $\mathbf f$ is the thermal noise whose statistics can also
be described by Eq.\ (\ref{white noise}) with $D=k_BT=1/\beta$. 
This TDGL equation is readily seen to be
exactly \eq(\ref{EOMNL3}) with $\mu_{\rm T}^0=\mu$.
Therefore, we conclude that  the ordered phase of 2D incompressible polar active fluids  has the same static (i.e., equal-time) scaling behaviors as the ordered phase of the 2D $XY$ model subject to the constraint $\mathbf{\nabla}\cdot\mathbf M=0$.

This mapping between a nonequilibrium active fluid model and 
a ``divergence-free'' $XY$ model allows us to investigate the fluctuations in our original active fluid model by  studying the partition function of the equilibrium model. 

To deal with the exact identity $\mathbf \nabla \cdot \mathbf u=0$, we use a trick familiar from the study of incompressible fluid mechanics: we introduce a ``streaming function"; i.e., a new scalar field $h(\br)$ such that
\beq
u_x = -v_0 \pp_y h
\sep
u_y = v_0 \pp_x h
\ .
\label{VTrans1}
\eeq
Because this construction guarantees that the incompressibility condition $\mathbf \nabla \cdot \mathbf u=0$ is automatically satisfied, there is no constraint on the field $h(\br)$.

\begin{figure}
 \begin{center}
\includegraphics[scale=.45]{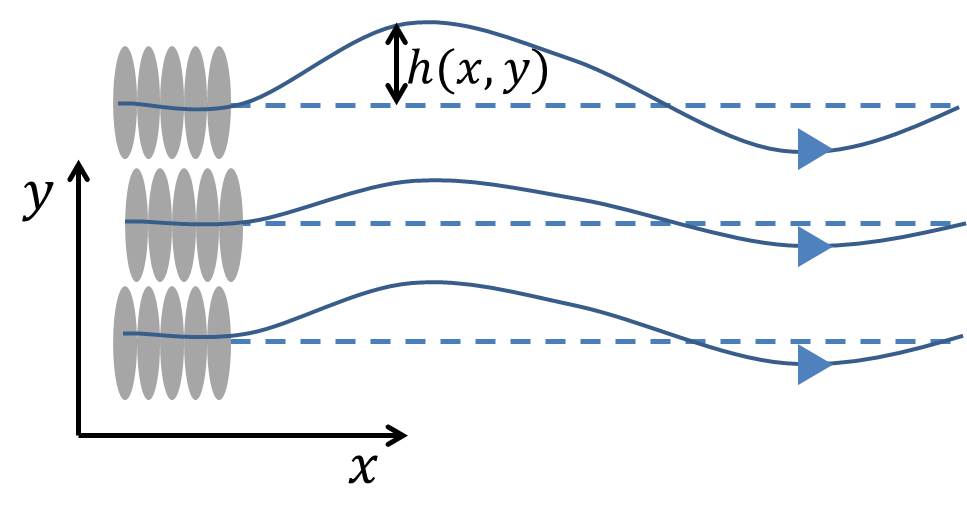}
 \end{center}
 \caption{$|$ {\bf Analogy between displacment field of the flow lines in 2D incompressible polar active fluids and that of 2D smectic layers.} In the case of 2D incompressible polar active fluids, the field $h(\br)$  is the vertical displacement of the flow lines (i.e., the solid lines) from the set of  parallel lines (i.e., the dotted lines) along $\hat{x}$ that would occur in the absence of fluctuations. For a defect-free 2D smectic, it likewise gives the vertical displacement of the smectic layers (i.e.,  the solid lines) from their reference positions (i.e., the dotted lines) at zero temperature.
 }
 \label{fig2}
 \end{figure}

The field $h(\br)$ has a simple 
interpretation 
as the displacement of the fluid flow lines from set of  parallel lines along $\hat{x}$ that would occur in the absence of fluctuations, as illustrated in  Fig. \ref{fig2}. (We thank Pawel Romanczuk for pointing out this pictorial interpretation to us.) 
This fact, which  is explained in more detail in  ``Methods", is a consequence of the fact that,
as in conventional 2D fluid mechanics, contours of the streaming function $h(\br)$ are flow lines.

This picture of a set of lines that ``wants" to be parallel being displaced by a fluctuation $h(\br)$ looks
very much like a  2D
smectic liquid crystal (i.e., ``soap"), for which the layers are actually 1D 
fluid stripes.
\newline

\noindent {\bf 2D smectic and KPZ models.} 
This resemblance between our system and a 2D  smectic is not purely visual. Indeed, making  the  substitution (\ref{VTrans1}), the Hamiltonian  (\ref{Hxy2}) becomes (ignoring irrelevant terms like $(\pp_x\pp_y h)^2$, which is irrelevant compared to $(\partial^2_x h)^2$ because $y$-derivatives are less relevant than $x$-derivatives):
\beq
H_s=\frac{1}{2} \int d^2r \left[B\left(\partial_y h-{(\partial_x h)^2\over 2}\right)^2+ K(\partial^2_x h)^2\right]
\ ,\label{Smec1}
\eeq
where $B=2\alpha v_0^2$ and $K=\mu v_0^2$. This Hamiltonian is exactly the Hamiltonian for the  dislocation-free 2D smectic model with $h(\br)$ in \eq(\ref{Smec1}) interpreted as the displacement field of the smectic layers, as also illustrated in Fig. \ref{fig2}.

The scaling behaviours of the dislocation-free 2D smectic  model are extremely non-trivial, since the ``critical dimension" $d_{\rm c}$ below which a purely harmonic description of these systems breaks down is $d_{\rm c}=3$\cite{GP}. 
Fortunately, these non-trivial scaling behaviours are known, thanks to 
an ingenious further mapping \cite{Golubo1,Golubo2} of this problem onto  the 1+1-dimensional KPZ 
equation \cite{KPZ}, which is a model for interface growth or erosion (e.g., ``sandblasting"). In 
this mapping, which connects the equal-time correlation functions of the 2D smectic to the 
KPZ equation, the $y$-coordinate in the smectic is mapped onto time $t$ in 
the 
KPZ equation with $h(x,t)$ the height of the ``surface" at position $x
$ and time $t$ above some reference height. As a result,  the 
dynamical exponent $z_{_{\rm KPZ}}$ of the  1+1-dimensional KPZ equation becomes the 
anisotropy exponent $\zeta$ of the 2D smectic. Since the scaling laws  of the 1+1-dimensional KPZ equation are known exactly \cite{KPZ}, those of the equal-time correlations of 
the 2D smectic can be obtained as well.

This gives \cite{Golubo1,Golubo2} $\zeta=3/2$ and $\chi_{_h}=1/2$ as the exponents for the 2D smectic, where $\chi_{_h}$ gives the scaling of the smectic layer displacement field
$h(\br)$ with spatial coordinate $x$. Given the streaming function relation (\ref{VTrans1}) between $h(\br)$ and $\bu(\br)$, we see that the scaling exponent $\chi_{_y}$ for $u_y$ is just
$\chi_{_y}=\chi_{_h}-1=-1/2$ and  that the scaling exponent $\chi_{_x}$ for $u_x$ is just
$\chi_{_x}=\chi_{_y}+1-\zeta=-1$.  Note that these exponents are different from those for 
compressible polar active fluids\cite{TT1, TT2, TT3,TT4,TT5} where $\zeta=5/3$  and $\chi_y=-1/5$.
(Note that our convention here ($q_y\sim q_x^\zeta$) is the inverse of that ($q_x\sim q_y^\zeta$) used in references \cite{TT1, TT2, TT3,TT4,TT5}.)

The fact that both of the scaling exponents $\chi_y$ and $\chi_x$ are less than zero implies that both  $u_y$ and  
$u_x$ fluctuations remain finite as system size $L\rightarrow\infty$; this, in turn, implies that the system  has 
long-ranged orientational order  since $\la |\bv(\br, t)-\bv(\br',t)|^2\ra$ remains finite as $|r-r'|\rightarrow \infty$.
That is, the ordered state is stable against fluctuations, at least for sufficiently small noise $D$.

The velocity correlation function can be calculated through
the connection between $\bu$ and $h$. Using the aforementioned connection between 2D smectics and the 1+1-dimensional KPZ equation, the  equal time layer displacement 
correlation function takes the form \cite{Golubo1,Golubo2}:
\begin{eqnarray}
C_h(\br-\br^\prime) &\equiv& \left<[h(\br,t)-h(\br^\prime, t)]^2\right>\nonumber\\&=&  B|x-x^\prime | \Psi(\kappa)
\label{h real space scaling} .
\end{eqnarray}
where we define the scaling variable 
$\kappa\equiv{X\over Y^{2/3}}$, with $X \equiv |x-x'|/\xi_x$, and $Y\equiv |y-y'|/\xi_y$,
and the non-universal constant $B$ is an overall multiplicative 
factor;
estimates of the non-universal nonlinear lengths $\xi_{x,y}$ are given in ``Methods". 

The limiting behaviors of the universal scaling function   $\Psi$ have been studied numerically previously \cite{tang,frey96}. Here, we use the most accurate
version currently known (www-m5.ma.tum.de/KPZ) \cite{ExactKPZ1,ExactKPZ2}:
\beq
\Psi(\kappa) \approx \left\{
\begin{array}{ll}
\Psi(\kappa) \approx c_1+e^{-\Phi_{h}(\kappa)} \ ,
 \ , & \kappa\gg 1 
\\\\
\kappa + {c_2\over \kappa} \ , & \kappa \ll  1
\ ,
\label{psih}
\end{array}
\right.
\eeq
where  for $\kappa\gg 1$, 
\beq
\label{Phi}
\Phi_{h}(\kappa)= c\kappa^3 +\cO(\kappa) \ .
\eeq
Here, the constants $c$ and $c_{1,2}$ are all universal and are given by $c\approx 0.295$, $c_1\approx 1.843465$, and $c_2\approx 1.060...$ \cite{ExactKPZ1,ExactKPZ2}.

Rewriting the velocity correlation function (\ref{LC_scaling})  in terms of the fluctuation $\bu$ using (\ref{Expension1}) gives 
\beqn
&&\la |\bv(\br,t)-\bv(\br', t)|^2\ra\nonumber\\
&=&C_0-2\la u_y(\br,t) u_y(\br',t) \ra-2\la u_x(\br,t) u_x(\br',t) \ra\,,~
\label{Vuu}
\eeqn
where $C_0=2\la u^2(\br,t)\ra$ is finite, and the two correlation functions on the right hand side of the equality are just the derivatives of the layer displacement correlation function:
\beqn
\la u_y(\br,t) u_y(\br',t) \ra &= -{ v_0^2\over 2}\pp_x\pp_{x'} C_h(\br-\br^\prime)\,,
\label{cy}\\
\la u_x(\br,t) u_x(\br',t) \ra &= -{v_0^2\over 2}\pp_y\pp_{y'} C_h(\br-\br^\prime)\,.
\label{cx}
\eeqn
To derive (\ref{cy}, \ref{cx}) we use (\ref{VTrans1}) and the definition of $C_h$ (i.e., the first equality of formula (\ref{h real space scaling})).

Inserting 
(\ref{cy}, \ref{cx}) into (\ref{Vuu}) and using the asymptotic forms (\ref{h real space scaling}, \ref{psih}) for $C_h$,
we obtain (as explained in more detail in  ``Methods") the asymptotic form of the velocity correlation function given by
(\ref{LC_scaling}). We can also obtain the Fourier transformed equal time correlation functions; these are given in  ``Methods".
\newline

\noindent{\bf Discussion}
\newline
We formulate a universal equation of motion describing the ordered phase of 2D
incompressible polar active fluids. 
After using renormalization group analysis to identify the relevant non-linearities of this model,
we perform  a series of mathematical transformation which map our model to three other interesting,  but seemingly unrelated,
models. Specifically, we  make heretofore unanticipated
connections between
four seemingly unrelated systems:
the ordered phase of 2D incompressible polar active fluids, the ordered phase of the divergence-free 2D XY model,  
dislocation-free 2D smectics, 
 and growing one-dimensional interfaces.
Through this connection, we show that 2D incompressible  polar active fluids
spontaneously break continuous rotational invariance (which their equilibrium 
counterparts (i.e., ordinary 
 divergenceful
ferromagnets) cannot
do), and
obtain the exact scaling behavior of the equal-time velocity correlation function of
the original model. Because this mapping
only involves equal-time correlations, 
the dynamical
scaling of  the original model is currently unknown.
We hope to determine this scaling in further work. 
\newline

\noindent {\bf Methods}
\newline
\noindent {\bf Linear theory.}
In this section we give the details of  the derivation of the linearized theory of incompressible polar active fluids. We begin with the linearized equation of motion, obtained by expanding Eq. (6) of the main text to linear order in the fluctuation $\bu$ of the velocity around its mean value $v_0\hat{x}$: 
\begin{widetext}
\beqn
\pp_t u_m = -\pp_m P -2\alpha u_x\delta_{mx} - \lambda^0_1v_0(\pp_x u_m)
- \lambda^0_4v_0^3\delta_{xm}(\pp_x u_x)
 +\mu^0_{\rm T} \nabla^2 u_m
+  \mu_2^0v_0^2 \pp_x^2u_m
+ f_m
\ ,\label{EOMlinSM}
\eeqn
\end{widetext}
where the superscript ``0'' means that the $v$-dependent coefficients are
evaluated at $v=v_0$, and we define the ``longitudinal mass"
$\alpha\equiv -{v_0\over 2}\left({dU(v)\over dv}\right)_{v=v_0}$.

Our goal now is to determine the scaling of the fluctuations $\bu$ of the velocity
with length and time scales, and to determine the relative scaling of the two
Cartesian components $x$ and $y$ of position with each other, and with time $t$. That is,
in the language of hydrodynamics, we seek the ``roughness exponents" $\chi_{x,y}$, the
anisotropy exponent $\zeta$, and the dynamical exponent $z$ characterizing respectively the scaling of: velocity fluctuations $u_{x,y}$, ``transverse" (i.e., perpendicular to the direction of flock motion) position $y$, and time $t$ with ``longitudinal" (i.e., parallel to the direction of flock motion) position $x$. Knowing this scaling (in particular, $\chi_{x,y}$) allows us to answer the most important question about this system: is the ordered state actually stable against fluctuations?

 To obtain this scaling in the linear theory, we begin by calculating the fluctuations of $\bu$
predicted by that theory.  Since the two components of $\bu$ are not
independent,  but, rather,  locked to each other by the incompressibility condition $\nabla\cdot\bv=0$, it is only necessary to
calculate one of them. We choose to focus on the $y$-component,  which can be calculated by first
spatio-temporally Fourier transforming (\ref{EOMlinSM}), and then acting on both sides with the
transverse projection operator $P_{lm}(\bq)=\delta_{lm}-q_lq_m/q^2$ which projects
orthogonal to the spatial wavevector $\bq$.
 The component $\ell=y$
of the resultant equation then gives
\begin{widetext}
\beqn
-i(\omega- \lambda^0_1v_0q_x)u_y(\bq,\omega)=(2\alpha
+ i\lambda^0_4v_0^3q_x){q_xq_y\over q^2}u_x(\bq, \omega) -\Gamma(\bq) u_y(\bq, \omega)+P_{ym} f_m(\bq, \omega)
\ ,\label{EOMFTSM}
\eeqn
\end{widetext}
where we define
\begin{eqnarray}
\Gamma(\bq)\equiv\mu_{\rm T}^0q^2+\mu_2^0v_0^2q_x^2=\mu q_x^2+\mu_{\rm T}^0q_y^2\, ,\label{Gamma1SM}
\end{eqnarray}
with $\mu\equiv \mu_{\rm T}^0+\mu_2^0v_0^2$.

We can eliminate $u_x$ from (\ref{EOMFTSM}) using the incompressibility condition
$\nabla\cdot\bv=0$, which implies, in Fourier space, $q_xu_x=-q_yu_y$.
Solving  the resultant linear algebraic equation for
$u_y(\bq,\omega)$ in terms of $f_m(\bq, \omega)$ gives
\beqn
u_y(\bq, \omega)={P_{ym}(\bq)f_m(\bq, \omega)\over -i\left[\omega-c(\hat{\bq})q\right]+\Gamma(\bq)+2\alpha\left({q_y\over q}\right)^2} \,,
\label{usolvSM}
\eeqn
where we define the direction-dependent ``sound speed"
\beqn
c(\hat{\bq})\equiv\lambda_1^0v_0{q_x\over q}+\lambda^0_4v_0^3{q_y^2q_x\over q^3}\,.
\eeqn

Using Eq.~(\ref{EOMFTSM}), we can obtain $\la |u_y(\bq, \omega)|^2\ra$ from the known correlations of the
random force $\bff$ (i.e., formula (4) in the main text). Integrating the result over all frequencies $\omega$,  and dividing by $2\pi$, gives the equal time, spatially Fourier transformed velocity autocorrelation:
\begin{eqnarray}
 \langle  |u_y(\bq, t)|^2\rangle={Dq_x^2\over 2\alpha q_y^2+\Gamma (\bq)q^2}\approx{Dq_x^2\over 2\alpha q_y^2+\mu q_x^4}
 \ .
 \label{Uu3SM}
\end{eqnarray}
where
the second, approximate equality applies for all $\bq\rightarrow \bf{0}$.  
This can be seen by noting that, for $q_y\gg q_x^2$ and $\bq\rightarrow \bf{0}$,
$q_y^2\gg \Gamma(\bq)q^2$, while for $q_y\lesssim q_x^2$ and $\bq\rightarrow \bf{0}$,  $
\Gamma(\bq)q^2\approx\mu q_x^4$. Hence, in both cases, (which together cover all
possible ranges of $\bq$ for  $\bq\rightarrow \bf{0}$), the approximation $2\alpha q_y^2+
\Gamma (\bq)q^2\approx  2\alpha q_y^2+\mu q_x^4$ is valid.

 Equation (\ref{Uu3SM}) implies that fluctuations  diverge most rapidly as $\bq\rightarrow \bf{0}$  if $\bq$ is taken to zero along a locus in the $\bq$ plane that obeys $q_y\lesssim q_x^2$;  along such a locus,  asymptotically, $\langle  |u_y(\bq, t)|^2\rangle\propto{1\over q^2}$. In contrast, along all other locii, i.e., those for which $q_y\gg q_x^2$,  $\langle  |u_y(\bq, t)|^2\rangle\propto{q_x^2\over q_y^2}\ll{1\over q^2}$. In this sense, one can say that the regime $q_y\lesssim q_x^2$ shows the largest fluctuations at small $\bq$; this implies the anisotropy exponent $\zeta=2$.

We can get the dynamical exponent
$z$ predicted by the linear theory by inspection of (\ref{usolvSM}), although some
care is required. The form of the first term in the denominator might suggest
$\omega\propto q$, which would imply $z=1$. However, the propagating
$c(\hat{\bq})q$ term in this expression does not appear in our final
expression (\ref{Uu3SM}) for the fluctuations; rather, these are controlled entirely
by the damping
%
%
term $\Gamma(\bq)+2\alpha\left({q_y\over q}\right)^2$.
Balancing $\omega$ against that term in the dominant regime of wavevector  $q_y\sim q_x^2$ gives $\omega\propto q_x^2$, which implies $z=2$.

Now we seek $\chi_y$, which determines whether or not the ordered state is stable against fluctuations in an arbitrarily large system. This can be obtained by looking at the real space fluctuations $\la u_y^2(\br, t) \ra=\int_{{q_x}\gtrsim{1\over L}}{d^2q\over (2\pi)^2} \langle|u_y(\bq, t)|^2\rangle$, where $L$ is the lateral extent of the system in the $x$-direction (its extent in the $y$-direction is taken for the purposes of this argument to be infinite). Using (\ref{Uu3SM}), this integral is readily seen to converge in the infra-red, and, hence, as system size $L\rightarrow\infty$. Since the integral is finite, and proportional to the noise strength $D$, it is clear that, for sufficiently small $D$, the transverse $u_y$ fluctuations in real space can be made small enough that long-ranged orientational order, and, hence, a non-zero $\langle\bv(\br,t)\rangle$, is preserved in the presence of fluctuations; the ordered state is stable against fluctuations for sufficiently small noise strength $D$.

The exponent $\chi_y$ can be obtained by looking at the departure $\delta u_y^2$ of the
 $u_y$ fluctuations from their infinite system limit: 
 $\delta u_y^2\equiv \la u_y^2(\br, t) \ra|_{ L=\infty}-\la u_y^2(\br, t)\ra|_L=\int_{{q_x}\lesssim{1\over L}}{d^2q\over
(2\pi)^2} \langle|u_y(\bq, t)|^2\rangle$;
we define the ``roughness exponent" $\chi_y$ by the way
this quantity scales with system size $L$: $\delta u_y^2\propto L^{2\chi_y}$. 
Note that this
definition of $\chi_y$ requires $\chi_y<0$, since it depends on the existence of an ordered state, which necessarily implies that the
velocity fluctuations $\delta u_y^2$ do not diverge as $L\rightarrow\infty$.
If $\la u_y^2(\br, t) \ra|_{L=\infty}$ is not finite, one can obtain $\chi_{y}$
by performing exactly the type of scaling argument outlined here directly on $\la u_y^2(\br, t) \ra|_L$ itself.

Approximating (\ref{Uu3SM}) for the
dominant regime of wavevector $q_y\sim q_x^2$, and changing  variables in the integral from
$q_{x,y}$ to $Q_{x,y}$ according to $q_x\equiv {Q_x\over L}$, $q_y\equiv {Q_y\over L^2}$
 shows that $\delta u_y^2\propto L^{-1}$, and hence $\chi_y=-{1\over 2}$.

 Note also that the fluctuations of $u_x$ are much smaller than those of $u_y$. This can be seen by using the incompressibility condition, which implies, in Fourier space, $u_x=-{q_yu_y\over q_x}$, which implies
 \begin{eqnarray}
 \langle  |u_x(\bq, t)|^2\rangle={Dq_y^2\over 2\alpha q_y^2+\Gamma (\bq)q^2}\approx{Dq_y^2\over 2\alpha q_y^2+\mu q_x^4}
 \ ,
 \label{UxSM}
\end{eqnarray}
which is clearly finite as $\bq\rightarrow\bf{0}$ along any locus; indeed, it is bounded above by ${D\over 2\alpha}$.

We can calculate a roughness exponent $\chi_{_x}$ for $u_x$ for the linear theory from this result
 exactly as we calculate the roughness exponent $\chi_y$ for $u_y$; we find $\chi_{_x}=1-\zeta+ \chi_y=-{3\over 2}$. We shall see in the next section that the first line of this equality also holds
 in the full non-linear theory, even though the values of the exponents $\chi_{_x}$, $\zeta$, and $\chi_y$ all change.

 The fact that $u_x$ has much smaller fluctuations than $u_y$ means that we have to work to higher order in $u_y$ than in $u_x$ when we treat the non-linear theory, as we do in next section.
\newline

\noindent{\bf Mapping to an equilibrium ``incompressible" magnet.}
We now go beyond the linear theory, and expand the full equation of motion (6) of the main text to higher order in $\bu$. We obtain
\begin{widetext}
\begin{eqnarray}
\pp_t u_m &=& -\pp_m P -2\alpha u_x\delta_{mx} - \lambda^0_1v_0\pp_x u_m
 +\mu^0_{\rm T} \nabla^2 u_m
+  \mu_2^0v_0^2 \pp_x^2u_m+f_m\nonumber
\\
&&-{\alpha\over v_0}\left[{u_y^3\over v_0}\delta_{my}+2u_xu_y\delta_{my}+u_y^2\delta_{mx}\right]
-\lambda^0_1u_y\pp_yu_{y}\delta_{my}
\ .\label{EOMNL1SM}
\end{eqnarray}
\end{widetext}

We keep terms that might naively
appear to be higher order in the small fluctuations (e.g., the $u_y^3\delta_{my}$ term relative to the  $u_xu_y\delta_{my}$ term) because, as we saw in the linearized theory, the two different components $u_{x,y}$ of $\bu$ scale differently at long length scales. Hence, it is not immediately obvious, e.g., which of the two terms just mentioned is actually most important at long distances. We therefore, for now, keep them both.
For essentially the same reason, it is not obvious whether $u_y^2\delta_{mx}$ or $u_y^3\delta_{my}$ is more important, so we shall for now keep both of these terms as well.

On the other hand, it is immediately obvious that a term like, e.g.,
$u_xu_y^2 \delta_{mx}$
is less relevant than
$u_y^2 \delta_{mx}$, since, whatever the relative scaling of $u_x$ and $u_y$,
$u_xu_y^2 \delta_{mx}$ is much smaller at large distances than
$u_y^2 \delta_{mx}$, since $u_x$ is.

Likewise, we drop the term ${1\over 2}\left(d\lambda_1 \over dv\right)_{v=v_0} u_y^2\pp_xu_ y\delta_{my}$,
since it is manifestly smaller, by one $\pp_x$, than the $u_y^3\delta_{my}$ term already displayed explicitly in (\ref{EOMNL1SM}).

This sort of reasoning guides us very quickly to the reduced model (\ref{EOMNL1SM}). As explained in the main text, acting on both sides of (\ref{EOMNL1SM})  with the transverse projection operator $P_{lm}(\bq)=\delta_{lm}-q_lq_m/q^2$ which projects orthogonal to the spatial wavevector $\bq$  eliminates the pressure term. Then taking the $l=y$
component of the resulting equation gives (11) of the main text, which we now use to calculate the rescaled coefficients.

To do this, we must also determine how the projection operators $P_{yx}$ and $P_{yy}$ rescale upon the rescalings (i.e., (12) of the main text).
Since in the linear theory
(see, e.g.,  the $u_y$--$u_y$ correlation function (\ref{Uu3SM}))
fluctuations are dominated
by the regime $q_y \lesssim q_x^2$, it follows that
$P_{yx}(\bq)=-{q_xq_y\over q^2}\approx -q_y/q_x\ll1$ and  $P_{yy}(\bq)=1-{q_y^2\over q^2}\approx 1$.
This implies that these  rescale according to
\beqn
&&P_{yx}(\bq) \mapsto  e^{\left(1-\zeta\right)\ell} P_{yx}(\bq)
\ , \
P_{yy}(\bq) \mapsto  P_{yy}(\bq) \ .
\label{PrescaleSM}
\eeqn

Performing the rescalings
(12-14) of the main text, and  (\ref{PrescaleSM}) above on the equation of motion (11) of the main text, we obtain, from the rescalings of first three 
(i.e., the linear) terms on the right hand side the following rescalings of the  parameters:
\beqn
\label{linrsSM}
&&v_1 \mapsto  e^{\left(z-1\right)\ell} v_1
\ , \ \mu\mapsto  e^{\left(z-2\right)\ell} \mu  \ ,
\ \nonumber\\
&& \mu_{\rm T}^0 \mapsto  e^{\left(z-2\zeta\right)\ell} \mu_{\rm T}^0
\ ,
\eeqn
and
\beqn
\label{linalphaSM}
\alpha \mapsto  e^{\left(z-2\zeta+2\right)\ell} \alpha
\ .
\eeqn
Note that the $\Gamma(\bq)$ term in (11) of the main text involves two parameters ($\mu$ and $\mu_{\rm T}^0$); hence, we get the rescalings of both of these parameters from this term.

Similarly, looking at the rescaling of the non-linear terms proportional to $u_y^2$ and $u_y^3$,  respectively, we obtain the rescalings:
\beqn
\label{nlalphaSM}
&&{\alpha\over v_0} \mapsto  e^{\left(z+\chi_y-\zeta+1\right)\ell} {\alpha\over v_0}
\ , \
{\alpha\over v^2_0} \mapsto  e^{\left(z+2\chi_y\right)\ell} {\alpha\over v^2_0}
\ .
\eeqn
We recover the first of these by looking at the rescaling of the non-linear term proportional to $u_xu_y$ as well.

We note that  the two rescalings (\ref{nlalphaSM})
are both consistent with
 (\ref{linalphaSM}) if we rescale
$v_0$ according to
\beqn
\label{v0rescSM}
v_0 \mapsto  e^{\left(1-\zeta-\chi_y\right)\ell} v_0
\ .
\eeqn

By power counting on the $u_y\pp_y u_y$ term, we obtain the rescaling of $\lambda_1^0$:
\beqn
\lambda_1^0 \mapsto e^{\left(z+\chi_y-\zeta\right)\ell}\lambda_1^0
\ .
\label{lamrescSM}
\eeqn

Finally, by looking at the rescaling of the noise correlations (i.e., (4) of the main text), we obtain the scaling of the noise strength $D$:
\beqn
&&D \mapsto  e^{\left(z-2\chi_y-\zeta-1\right)\ell} D
\ .
\label{DrescSM}
\eeqn

We now use the standard renormalization group logic to assess the importance of the non-linear terms in (11) of the main text. This logic is to choose  the rescaling exponents $z$, $\zeta$, and $\chi_y$ so as to keep the size of the fluctuations in the field $\bu$ fixed upon rescaling. Since, as we 
saw in our treatment of the linearized theory (in particular, \eq(\ref{Uu3SM})), that size is controlled by  three parameters: the ``longitudinal mass"  $\alpha$, the damping coefficient $\mu$, and the noise strength $D$, the choice of $z$, $\zeta$, and $\chi_y$ that keeps these fixed will clearly accomplish this.
From the rescalings (\ref{linrsSM}), (\ref{linalphaSM}), and (\ref{DrescSM}), this leads to three simple linear
equations in the three unknown exponents $z$, $\zeta$, and $\chi_y$; solving these, we find the values of these exponents in the linearized theory:
\beqn
\zeta_{_{\rm lin}}=z_{_{\rm lin}}=2,\  \ \chi_{_{y\rm lin}}=-1/2,\ \ \chi_{_{x\rm lin}}=-3/2 \,
\label{linSM}
\eeqn
which, unsurprisingly, are the linearized exponents we found earlier.

With these exponents in hand, we can now assess the importance of the non-linear terms in (11) of the main text at long length scales, simply by looking at how their coefficients rescale. (We don't have to worry about the size of the actual non-linear terms themselves changing upon rescaling, because we have chosen the rescalings to keep them constant  in the linear theory.)
The mass $\alpha$, of course, is kept fixed. Inserting the linearized exponents (\ref{linSM}) into the rescaling relation (\ref{v0rescSM}) for $v_0$, we see that
\beqn
\label{v0resc2SM}
v_0 \mapsto  e^{-{\ell\over 2}} v_0
\ .
\eeqn
Since $v_0$ appears in the denominator of all three of the non-linear terms associated with $\alpha$, and $\alpha$ itself is fixed, this implies that all three of those terms are ``relevant", in the renormalization group sense of growing larger as we go to longer wavelengths (i.e., as $\ell$ grows).
As usual in the RG, this implies that these terms ultimately alter the scaling behavior of the system at sufficiently long distances. In particular, the exponents
$z$, $\zeta$, and $\chi_{x,y}$ change from their values (\ref{linSM}) predicted by the linear theory.

The same is not true of the $\lambda^0_1$ non-linearity,
however, because it is irrelevant; that is, it gets smaller upon renormalization. This follows from inserting 
the linearized exponents (\ref{linSM}) into the rescaling relation (\ref{lamrescSM}) for $\lambda^0_1$, which gives
\beqn
\label{lamresc2SM}
\lambda^0_1 \mapsto  e^{-{\ell\over 2}} \lambda^0_1
\ ,
\eeqn
which shows clearly that $\lambda_1^0$ vanishes as $\ell\rightarrow\infty$; that is, in the long-wavelength limit.

Since $\lambda_1^0$ was the only remaining non-linearity associated with the $\lambda$ terms in our original equation of motion (\ref{EOMNL1SM}),  we can accurately treat the full, long distance behavior of this problem by leaving out all of those non-linear terms.

Doing so
reduces the equation of motion
(\ref{EOMNL1SM}) to
\begin{eqnarray}
\pp_t u_m =&& -\lambda^0_1v_0\pp_xu_m-\pp_m P -2\alpha \left(u_x+{u_y^2\over 2v_0}\right)\delta_{xm} \nonumber\\
&&-{2\alpha\over v_0} \left(u_x+{u_y^2\over 2v_0}\right)u_y\delta_{ym} \nonumber\\
&&+  \mu \pp_x^2u_m+\mu_{\rm T}^0\pp_y^2u_m+f_m
\ .\label{EOMNL2}
\end{eqnarray}

Before proceeding to analyze this equation, we note  the differences between  the structure of this problem and that
for the compressible case. In the compressible problem, there is no
constraint analogous to the incompressibility condition relating $u_x$ and $u_y$. Hence, $u_x$
is free to relax quickly (to be precise, on a time scale ${1\over 2\alpha}$) to its local ``optimal" value,
which is readily seen to be $-{u_y^2\over 2v_0}$. Once this relaxation has occurred, all of the non-linearities associated with $\alpha$ drop out of that compressible problem, leaving the $\lambda$ non-linearities as the dominant ones. For a detailed discussion of the rather tricky analysis of the compressible problem that leads to this conclusion, see Ref. \cite{rean}. Here, in the incompressible problem, $u_x$ is, because of the incompressibility constraint,  not free to relax in such a way as to cancel out the $\alpha$ non-linearities, which, because they involve no spatial derivatives, wind up dominating the $\lambda$ non-linearities, which do involve spatial derivatives. In addition, the suppression of fluctuations by the incompressibility condition, which as we've already seen in the linear theory, makes the $\lambda$ non-linearities not only less relevant than the $\alpha$ ones, but actually irrelevant. Hence, we can drop them in this incompressible problem, leaving us with
\eq(\ref{EOMNL2}) as our equation of motion.

As one final simplification, we make a Galilean transformation to a ``pseudo-co-moving" co-ordinate system moving in the direction $\hat{{x}}$ of mean flock motion at speed $\lambda^0_1v_0$. 
Note that if the parameter $\lambda_1^0$ had been equal to $1$, this would be precisely the frame co-moving with the flock. The fact that it is not is a consequence of the lack of Galilean invariance in our problem.

This boost eliminates the ``convective'' term $\lambda^0_1v_0\pp_xu_m$ from the right hand side of (\ref{EOMNL2}), leaving us with our final simplified form for the  equation of motion:
\begin{eqnarray}
\pp_t u_m =&& -\pp_m P -2\alpha \left(u_x+{u_y^2\over 2v_0}\right)\delta_{xm} \nonumber\\
&&-{2\alpha\over v_0} \left(u_x+{u_y^2\over 2v_0}\right)u_y\delta_{ym} \nonumber\\
&&+  \mu \pp_x^2u_m+\mu_{\rm T}^0\pp_y^2u_m+f_m
\ .\label{EOMNL3SM}
\end{eqnarray}
which is just equation (15) of the main text.
\newline

\noindent{\bf Mapping of equilibrium ``incompressible" magnet to 2D smectic.}
We begin by demonstrating the pictorial interpretation of the  ``streaming function" introduced in the main text via
\beq
u_x = -v_0 \pp_y h
\sep
u_y = v_0 \pp_x h
\ .
\label{VTrans1SM}
\eeq
This implies that the streaming function $\phi$ for the full
velocity field $\bv(\br)=v_0\hat{x}+\bu$, defined via $v_x = \pp_y \phi
$, $v_y = -\pp_x \phi$, is given by
\beqn
\phi=v_0(y-h(\br))\,.
\label{phi1SM}
\eeqn

As in conventional 2d fluid mechanics, contours of the streaming function $\phi$ are flow lines.
 When the system is in its uniform steady state
 (i.e., $\bv=v_0\hat{x}$), these contour lines, defined via
\begin{eqnarray}
 \phi=nC,~~~~n=0,1,2,3...
 \label{contours1SM}
\end{eqnarray}
where $C$ is some arbitrary constant,
are a set of parallel, uniformly spaced lines given by $y_n=nC/v_0$.

Now let's ask what the
flow lines are if there are fluctuations in the velocity field: $\bv=v_0\hat{x}+\bu$.
Combining our expression for $\phi$ (\ref{phi1SM}) and the expression (\ref{contours1SM}) for the flow lines, we see that the positions of the flow lines are now given by
\begin{eqnarray}
y_n=nC/v_0+h,~~~~n=0,1,2,3...\,,
 \label{contours2SM}
\end{eqnarray}
which shows that $h(\br)$
can be interpreted as the local displacement of the flow lines from their positions in the ground state configuration.

This picture of a set of lines that ``wants" to be parallel being displaced by a fluctuation $h(\br)$ looks
very much like a 2D smectic liquid crystal (i.e., ``soap"), for which the layers are actually one-dimensional fluid stripes.

This resemblance between our system and a 2D smectic is not purely visual. Indeed, making  the  substitution (\ref{VTrans1SM}), the Hamiltonian  (17) of the main text becomes (ignoring irrelevant terms like $(\pp_x\pp_y h)^2$, which is irrelevant compared to $(\partial^2_x h)^2$ because $y$-derivatives are less relevant than $x$-derivatives)
\beq
H_{\rm s}=\frac{1}{2} \int d^2r \left[B\left(\partial_y h-{(\partial_x h)^2\over 2}\right)^2+ K(\partial^2_x h)^2\right]
\ ,\label{Smec1SM}
\eeq
where $B=2\alpha v_0^2$ and $K=\mu v_0^2$. This Hamiltonian is exactly the Hamiltonian for the equilibrium 2D smectic model with $h$ in \eq(\ref{Smec1SM}) interpreted as the displacement field of the smectic layers. For the equilibrium 2D smectic the partition function is
\beq
\label{eq:ZsSM}
Z_{\rm s}= \int D[h]\ee^{-H_{\rm s}/k_BT}
\ ,
\eeq
where it should be noted that there is no constraint on the functional integral over $h(\br)$ in this expression, since, as noted earlier, $h(\br)$ is unconstrained.

Since the variable transformation Eq.\ (\ref{VTrans1SM}) is linear,  the partition functions  for the smectic: $Z_{\rm s}$ (\eq (\ref{eq:ZsSM})) and that for the constrained $XY$ model:
\beq
\label{eq:ZXYSM}
Z_{\rm _{XY}}= \int D[u_x]D[u_y]\delta(\nabla\cdot\mathbf u=0)\ee^{-H_{\rm _{XY}}/k_BT}
\ ,
\eeq
are the same up to a constant Jacobian factor, which changes none of the statistics.

To summarize what we have learned so far: we have successfully mapped the model for the ordered phase of an incompressible polar active
fluid onto the ordered phase of the equilibrium 2D $XY$ model with the constraint $\nabla\cdot\mathbf u=0$,
which in turn we have mapped onto   the standard equilibrium 2D smectic model \cite{Kashuba} 
The scaling
behaviors of the former can therefore be obtained by studying the latter.
Note that the connection between our problem and the dipolar magnet, which was studied in \cite{Kashuba}, is that the long-ranged dipolar interaction in magnetic systems couples to, and therefore suppresses, the longitudinal component of the magnetization. See \cite{Kashuba} for more details.
\newline

\noindent{\bf Mapping the 2D smectic to the one-dimensional KPZ equation.}
Fortunately, the scaling behaviours  of the equilibrium 2D smectic  model are known, thanks to 
an ingenious further mapping \cite{Golubo1,Golubo2} of this problem onto  the 1+1-dimensional KPZ 
equation \cite{KPZ}, which is a model for interface growth or erosion (e.g., ``sandblasting"). In 
this mapping, which connects the equal-time correlation functions of the 2D smectic to the 
1+1-dimensional KPZ equation, the $y$-coordinate in the smectic is mapped onto time $t$ in 
the 1+1-dimensional KPZ equation with $h(x,t)$ the height of the "surface" at position $x
$ and time $t$ above some reference height. As a result,  the  
dynamical exponent $z_{\rm _{KPZ}}$ of the  1+1-dimensional KPZ equation becomes the 
anisotropy exponent $\zeta$ of the 2d smectic. Since the scaling laws  of the 1+1-
dimensional KPZ equation are known exactly \cite{KPZ}, those of the equal-time correlations of 
the 2D smectic can be obtained as well.

This gives \cite{Golubo1,Golubo2} $\zeta=3/2$ and $\chi_{_h}=1/2$ as the exponents for the 2D smectic, where $\chi_{_h}$ gives the scaling of the smectic layer displacement field
$h$ with spatial coordinate $x$. Given the streaming function relation (\ref{VTrans1SM}) between $h$ and $\bu$, we see that the scaling exponent $\chi_{_y}$ for $u_y$ is just
$\chi_{_y}=\chi_{_h}-1=-1/2$ and  that the scaling exponent $\chi_{_x}$ for $u_x$ is just
$\chi_{_x}=\chi_{_y}+1-\zeta=-1$. 

The fact that both of the scaling exponents $\chi_y$ and $\chi_x$ are less than zero implies that both  $u_y$ and  $u_x$ fluctuations remain finite as system size $L\rightarrow\infty$; this, in turn, implies that the system  has long-ranged orientational order. That is, the ordered state is stable against fluctuations, at least for sufficiently small noise $D$.

The velocity correlation function can be calculated through
the connection between $\bu$ and $h$. Using the aforementioned connection between 2D smectics and the $1+1$-dimensional KPZ equation, the  layer displacement correlation 
function takes the form \cite{Golubo1, Golubo2}:
\begin{eqnarray}
C_h(\br-\br^\prime) &\equiv& \left<[h(\br,t)-h(\br^\prime, t)]^2\right>\nonumber\\&=& B|x-x^\prime |\Psi
(\kappa)
\label{h real space scalingSM} .
\end{eqnarray}
where the scaling variable $\kappa\equiv{X\over Y^{2/3}}$, $X =|x-x'|/\xi_x$, and $Y=|y-y'|/\xi_y$, 
$B$ is a non-universal overall multiplicative factor extracted from the scaling function,
and the non-universal nonlinear lengths $\xi_{x,y}$ are calculated in the next section. The universal scaling function   $\Psi$  has been numerically estimated (www-m5.ma.tum.de/KPZ) \cite{tang,frey96,ExactKPZ1,ExactKPZ2}:
\beq
\Psi(\kappa) \approx \left\{
\begin{array}{ll}
c_1+e^{-\Phi_{h}(\kappa)} 
 \ , & \kappa\gg 1 
\\\\
\kappa + {c_2\over \kappa} \ , & \kappa \ll  1
\ ,
\label{psihSM1}
\end{array}
\right.
\eeq
where  for $\kappa\gg 1$, 
\beq
\label{Phih}
\Phi_{\rm h}(\kappa)= c\kappa^3 +\cO(\kappa) \ .
\eeq
Here, the constants $c$ and $c_{1,2}$ are all universal and are given by $c\approx 0.295$, $c_1\approx 1.843465$,  and $c_2\approx 1.060...$ (www-m5.ma.tum.de/KPZ) \cite{ExactKPZ1,ExactKPZ2}. 
Only the lengths $\xi_{x,y}$, and the  overall multiplicative factor of $B$ in (\ref{h real space scalingSM}) are non-universal (i.e., system-dependent).

Rewriting the velocity correlation function (Eq. (1) of the main text) in terms of the fluctuations $\bu$ of the velocity from its mean value (as defined in Eq. (7) of 
the main text), we find 
\begin{eqnarray}
&&\la|\bv(\br,t)-\bv(\br', t)|^2\ra\nonumber\\
&=&C_0-2\la u_y(\br,t) u_y(\br',t) \ra-2\la u_x(\br,t) u_x(\br',t) \ra\,,~~~~\,\label{VuuSM}
\end{eqnarray}
where $C_0=2\la |\bu(\br,t)|^2\ra$  is finite.  The two correlation functions on the right hand side of the equality 
are just the derivatives of the layer displacement correlation function:
\beqn
\la u_y(\br,t) u_y(\br',t) \ra &=& -{v_0^2\over 2}\pp_x\pp_{x'} C_h(\br-\br')
\nonumber\\
&=&{Bv_0^2\over 2|x-x'|}\Psi_y(\kappa)\,,\label{cySM}
\\
\la u_x(\br,t) u_x(\br',t) \ra &=& -{v_0^2\over 2}\pp_y\pp_{y'} C_h(\br-\br')
\nonumber\\
&=&{Bv_0^2\xi_x^3\over 9\xi_y^2|x-x'|^2}\Psi_x(\kappa)\,.
\label{cxSM}
\eeqn
The velocity component scaling functions $\Psi_{x,y}$ can both be expressed in terms of the height scaling function $\Psi$, via

\beqn
\Psi_y(\kappa)&=& \kappa\left(2\Psi'+\kappa\Psi''\right)\,,
\label{psiySM}\\
\Psi_x(\kappa)&=&\kappa^4\left(5\Psi'+2\kappa\Psi''\right)\,.
\label{psixSM}
\eeqn
Using the asymptotic forms (\ref{psihSM1})  for the height scaling function $\Psi$ in (\ref{psiySM}) and (\ref{psixSM}), we obtain the asymptotic behaviors: 
\beqn
\Psi_y(\kappa) &\approx& \left\{
\begin{array}{ll}
2\kappa , &\, \,\,\kappa \ll 1 \ ,
\\\\
9c^2\kappa^6 e^{-\Phi_{h}(\kappa)} \ , &\, \,\,\kappa\gg  1 \ ,
\label{psiyasympSM}
\end{array}
\right.
\eeqn
\beqn
\Psi_x(\kappa) &\approx& \left\{
\begin{array}{ll}
-c_3\kappa^2, &\,\, \kappa \ll  1 \ ,
\\\\
18c^2\kappa^9e^{-\Phi_{h}(\kappa)} \ ,  &\,\, \kappa\gg  1 \ .
\label{psixasympSM}
\end{array}
\right.
\eeqn

Using these expressions (\ref{psiyasympSM}, \ref{psixasympSM}) for the scaling functions in the scaling  
expressions (\ref{cySM}, \ref{cxSM}) for the $u_x$ and $u_y$ correlation functions, and in turn using those 
in our expression (\ref{VuuSM}) for the velocity correlation function, we obtain 
\begin{eqnarray}
&& 
\left<|\bv(\br,t)-\bv(\br',t)|^2\right> \nonumber\\
&=&\left\{
\begin{array}{ll}
C_0-AY^{-2/3} \ , & \kappa\ll 1
\\
C_0-{9\over 2}c^2{A\over X} e^{-\Phi(\kappa)}\left[1+{4\over 9} \left({x-x'\over y-y'}\right)^2\right] \ , & \kappa\gg 1
\end{array}~~~
\right.\label{MLC_scaling}
\end{eqnarray}
where the non-universal constant $A$ is given by
\begin{eqnarray}
 &&A=2Bv_0^2\xi_x^{-1}\,. 
 \label{A1}
 \end{eqnarray}
 We've also defined a new universal function 
 \beq
 \Phi(\kappa)\equiv\Phi_{h}(\kappa)-6\ln(\kappa) \ ,
 \label{Phi}
 \eeq
which has the same limiting behaviour as $\Phi_{h}$, namely,
\beq
\label{Phi}
\Phi(\kappa)= c\kappa^3 +\cO(\kappa) \ .
\eeq
since the additive logarithm in (\ref{Phi}) is sub-dominant to the leading $\kappa^3$ term.
\newline

\noindent{\bf Calculation of the nonlinear lengths.}
The nonlinear lengths $\xi_{x,y}$ can be calculated most conveniently from the equilibrium 2D smectic model  (\ref{Smec1SM}). By definition,
$\xi_{x}$  and $\xi_{y}$ are the lengths along $x$ and $y$ beyond which the anharmonic terms in Eq. (\ref{Smec1SM})
become important. To determine these lengths, we treat the anharmonic terms perturbatively,
and calculate the lowest order correction to the harmonic terms. In a finite system of linear dimensions  $L_{x,y}$, this ``perturbative" correction will
indeed be perturbative (i.e., small) compared to the ``bare" values of the harmonic terms. However, they grow without bound with increasing $L_{x,y}$, and, hence, eventually cease to be small; that is, the perturbation theory
breaks down at large $L_{x,y}$. The values of $L_{x,y}$ above which the perturbation theory breaks down are the nonlinear lengths $\xi_{x,y}$.

\begin{figure}
\begin{center}
\includegraphics[scale=.25]{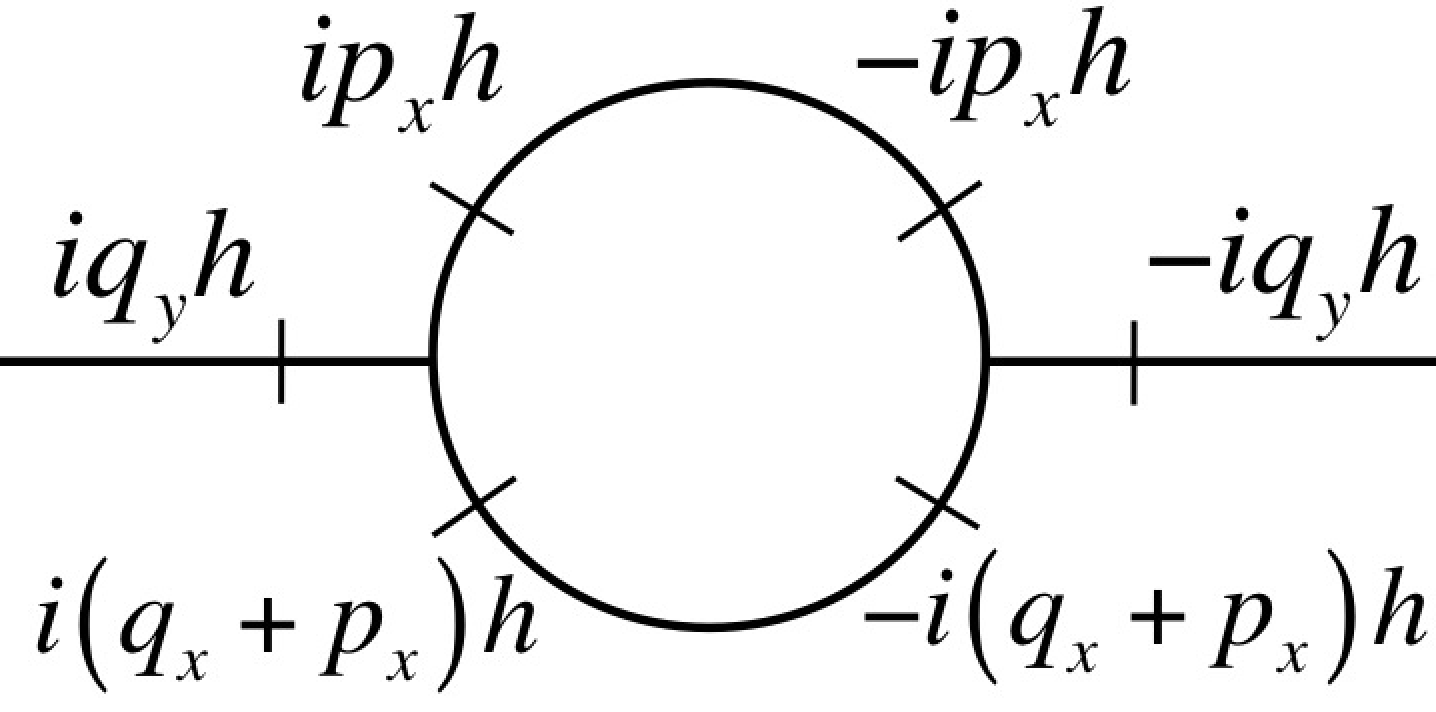}
\end{center}
\caption{$|${\bf The one-loop graphical correction to the compression modulus $B$ in the Hamiltonian (\ref{Smec1SM}).}
This arises from the combination of two cubic terms.
}
\label{fig3}
\end{figure}

Calculating 
 the lowest order correction to 
 the compression modulus $B$ (i.e., the coefficient of  $\left(\partial_y h\right)^2$ in the smectic Hamiltonian  (\ref{Smec1SM}))
 can be graphically represented by the Feynman diagram in Fig. \ref{fig3}. This leads to a correction to compression modulus:
\begin{eqnarray}
 \delta B&=&-{k_B TB^2\over (2\pi)^2}\int_{1\over L_x}^{\infty} dq_x\int_{-\infty}^{\infty} dq_y\ \ {q_x^4\over
                  \left(Bq_y^2+Kq_x^4\right)^2}\nonumber\\
             &=&-{k_BT\over 8\pi}\left(B\over K\right)^{3\over 2}\int^{\infty}_{1\over L_x}dq_x\ \ {1\over  
                    q_x^2}\nonumber\\
             &=&-{k_BT\over 8\pi}\left(B\over K\right)^{3\over 2}
                       L_x\,.
                       \end{eqnarray}
In this calculation, we have taken $L_y$, the system size 
along $y$, to be infinite. By the definition
of $\xi_x$, $|\delta B|= B$ for $L_x=\xi_x$, which gives 
\begin{eqnarray}
 \xi_x={8\pi B\over k_BT}\left(K\over B\right)^{3\over 2}
        ={4\sqrt{2}\pi v_0^2\mu^{3\over 2}\over D\alpha^{1\over 2}}\,,
\end{eqnarray} 
where in the second equality we have used the relations $B=2\alpha v_0^2$, $K=\mu v_0^2$, and $k_BT=D$ between the parameters of the smectic and those of the original incompressible active fluid.

Likewise, doing the same calculation for $L_y=\xi_y$, $L_x=\infty$,
we find
\begin{eqnarray}
 \xi_y={64\sqrt{2}\pi^2v_0^4\mu^{5\over 2}\over D^2\alpha^{1\over 2}}\,.
\end{eqnarray}
\newline

\noindent{\bf Fourier transformed correlation functions.}
The spatially Fourier transformed autocorrelations are also of interest. Fourier transforming (\ref{h real space scalingSM}) gives 
\beq
\la|h(\bq, t)|^2\ra=-{1\over 2}\int  dxdy e^{i(q_xx+q_yy)}|x|\Psi
\left({\left({|x| \over \xi_x} \right) \over
\left({|y|\over \xi_y}\right)^{2\over 3} }\right)\,.
\label{hq1SM}
\eeq
With the change of variables to new variables of integration $S$ and $W$ via $x\equiv {S\over q_x}$ and $y\equiv {W \xi_y\over (q_x\xi_x)^{3/2}}$, we immediately obtain
\beq
\la|h(\bq, t)|^2\ra=q_x^{-{7\over 2}}f(q_y/q_x^{3/2}) \,,
\label{hq2SM}
\eeq
with
\beq
f(\Theta)\equiv -{{\xi_y\over 2\xi_x^{3/2}}}\int  dSdW e^{i(S+\Theta W)}|S|\Psi
\left({|S|  \over
|W|^{2\over 3} }\right) \,.
\label{fhdefSM}
\eeq 
Combining \eq({\ref{hq2SM}) with the Fourier transform of the variable transformation (19) of the main text, we obtain the correlation functions for the ordered phase of the constrained equilibrium 2D $XY$ model, and hence, the ordered phase of incompressible active fluids:
\beqn
\la|u_y(\bq, t)|^2\ra&=&q_x^2\la|h(\bq, t)|^2\ra=q_x^{-{3\over 2}}f(q_y/q_x^{3/2})\label{Uuxc1SM}
\\
\la|u_x(\bq, t)|^2\ra&=&q_y^2\la|h(\bq, t)|^2\ra=q_y^2q_x^{-{7\over 2}}f(q_y/q_x^{3/2})  \nonumber\\&=&q_x^{-{1\over 2}}g(q_y/ q_x^{3/2}) \,,
\label{Uuyc1SM}
\eeqn
where $g(w)\equiv w^2 f(w)$. The limiting behaviors of the scaling functions $f(x)$ and $g(x)$ are:
$f(\xto)\rightarrow\rm{constant}\ne 0$, $f(x\rightarrow\infty)\propto x^{-7/3}$, $g(\xto)\propto x^2$, $g(x\rightarrow\infty)\propto x^{-1/3}$.
\newline

\noindent {\bf Data availability.} 
The data that support the findings of this study are available from any of the corresponding authors upon request.

\end{document}